\documentclass[10pt,journal,compsoc]{IEEEtran}
\usepackage{caption}
\usepackage{subcaption}
\usepackage{booktabs}
\usepackage{listings}
\usepackage{enumitem}
 \usepackage[most]{tcolorbox}
\usepackage{multirow}
\usepackage{graphicx}

\usepackage[normalem]{ulem}
\useunder{\uline}{\ul}{}
\usepackage{graphicx}
\usepackage{pifont}
\graphicspath{ {./fig/} }
\usepackage{url}
\usepackage{hyperref}
\usepackage{xspace}
\usepackage{amsmath,amssymb,amsfonts}

\newcommand{\tool}{\textsc{E-SPI}\xspace}

\newcommand{\wbz}[1]{\textcolor{black}{#1}}
\ifCLASSINFOpdf
\else
\fi
\begin{document}
%
\title{Enhancing Security Patch Identification by Capturing Structures in Commits}
\author{Bozhi~Wu,~Shangqing~Liu,~Ruitao~Feng,~Xiaofei~Xie,~Jingkai~Siow,~and~Shang-Wei~Lin
\IEEEcompsocitemizethanks{
\IEEEcompsocthanksitem S.~Liu is the corresponding author.
\IEEEcompsocthanksitem B.~Wu, S.~Liu, R.~Feng, J.~Siow and S. W.~Lin are with Nanyang Technological University, Singapore. \protect
E-mail: \{bozhi001, shangqin001, jingkai001\}@e.ntu.edu.sg, \{rtfeng,shang-wei.lin\}@ntu.edu.sg
\IEEEcompsocthanksitem X. Xie is with Singapore Management University, Singapore. E-mail: xiaofei.xfxie@gmail.com
}
}
\markboth{Journal of \LaTeX\ Class Files,~Vol.~14, No.~8, August~2015}%
{Shell \MakeLowercase{\textit{et al.}}: Bare Demo of IEEEtran.cls for IEEE Journals}
%



\IEEEtitleabstractindextext{%
\begin{abstract}
With the rapid increasing number of open source software (OSS), the majority of the software vulnerabilities in the open source components are fixed silently, which leads to the deployed software that integrated them being unable to get a timely update. Hence, it is critical to design a security patch identification system to ensure the security of the utilized software. However, most of the existing works for security patch identification just consider the changed code and the commit message of a commit as a flat sequence of tokens with simple neural networks to learn its semantics, while the structure information is ignored. To address these limitations, in this paper, we propose our well-designed approach \tool, which extracts the structure information hidden in a commit for effective identification. Specifically, it consists of the code change encoder to extract the syntactic of the changed code with the BiLSTM to learn the code representation and the message encoder to construct the dependency graph for the commit message with the graph neural network (GNN) to learn the message representation. We further enhance the code change encoder by embedding contextual information related to the changed code. To demonstrate the effectiveness of our approach, we conduct the extensive experiments against {six} state-of-the-art approaches on the existing dataset and from the real deployment environment. The experimental results confirm that our approach can significantly outperform current state-of-the-art baselines. 

\end{abstract}

}

\maketitle

\IEEEdisplaynontitleabstractindextext

%
\IEEEpeerreviewmaketitle

\IEEEraisesectionheading{\section{Introduction}}
\IEEEPARstart{R}ecently, a critical zero-day vulnerability named Log4Shell~\cite{Log4Shell} has attracted widespread attention in the security community. It exists in the widely used Java logging library Log4j~\cite{log4j}, which has been deployed by millions of Java applications. Specifically, due to this vulnerability can be exploited by attackers and leads to a remote code execution (RCE) when Log4j is used for logging, it was rated as ``very critical'' and received a CVSS~\cite{cvss} score of 10 (the highest level) . Although Log4Shell threats the security of many applications severely, however, due to the widespread attention, it has been resolved in a timely manner. On the contrary, most vulnerabilities are fixed silently, especially in the ``big code'' era, where open source software (OSS) has reached an unprecedented amount. According to the data from SourceClear~\cite{SourceClear2017}, 53\% of vulnerabilities in open source libraries are not publicly disclosed with CVEs. Hence, if the widely used open source components have unknown security risks, even if these risks are fixed in a timely manner by the developers of these components, the software that has already integrated them may not be updated in time, which causes substantial financial and social damages. Therefore, it is essential to identify these silent security patches.

Security patch identification is a crucial yet far from the settled problem, manual verification is a straightforward but unrealistic solution. According to the official report~\cite{Github2018} released by the largest global open source software (OSS) platform GitHub, 100 million projects have hosted on their platform in 2018. A huge amount of security related patches are committed daily for fixing, which makes the manual verification time-consuming and labor-intensive. Hence, automated security patch identification is an inevitable choice.

Conventional software analysis techniques that have been widely used in software vulnerability detection, such as static analysis~\cite{chess2004static} and dynamic analysis~\cite{ball1999concept} are not appropriate for security patch identification, because these techniques cannot be applied to the partial code (i.e., the changed code) in a commit. Furthermore, they also cannot support analysing the description of a commit (i.e., commit message), which is a form of natural text.

Inspired by the great success of machine learning and deep learning techniques in various fields ~\cite{wang2016cnn, bahdanau2014neural,arp2014drebin, wu2021android}, many learning-based approaches are proposed for automated security patch identification. For example, Zhou et al.~\cite{zhou2017automated} proposed to extract the features from the commit message and bug reports and further utilized a stacking of six machine learning classifiers such as random forest~\cite{breiman2001random} and SVM~\cite{noble2006support} to identify security issues. Another advanced deep learning based technique SPI~\cite{zhou2021spi} encoded both commit message and code changes with the BiLSTM encoder followed by the CNN layer to learn better representations. However, most existing works for patch identification only consider the changed code and commit message as a flat sequence of tokens, while ignoring the structure information hidden in the text. The structure information of the program has been widely demonstrated to improve various code-related tasks, such as software vulnerability identification~\cite{zhou2019devign}, source code summarization~\cite{liu2021retrievalaugmented}, and function name prediction~\cite{alon2018code2seq, alon2019code2vec}. Inspired by these works, Commit2vec~\cite{lozoya2021commit2vec} further extracted AST paths related to the changed code and fed them to the fully connected layer to learn the representation of the changed code to identify security-relevant commits. Commit2vec has achieved promising results, however, it only utilized the simple fully connected layer for learning, which has a limited learning capacity. Furthermore, the commit message is also ignored in utilization for the detection. 

To address the aforementioned challenges, in this paper, we propose our well-designed tool for security patch identification named \tool. Specifically, to capture the structure information of the commit, we design two separate encoders, AST-based code change encoder, which extracts the contextual AST paths related to the changed code with the BiLSTM encoder to learn the semantics of the changed code; graph-based commit message encoder, which constructs the commit message by the dependency graph with gated graph neural network (GGNN) to capture the token relations in the commit message. We further ensemble both encoders for the security patch identification. In addition, to capture the contextual information that related to the changed code, we propose to extract the AST paths that are not limited within the changed code (i.e., Within-Changes), but also include the paths related to the changed code (i.e., Within-Context) to enhance the code change representation produced by the code change encoder. Extensive experiments on the existing dataset and the real deployment environment have demonstrated the effectiveness of our approach. Specifically, \tool outperforms current state-of-the-art approaches significantly with {4.01\%} higher accuracy and {4.22\%} F1 score on the existing dataset, increases 6.03\% accuracy and 7.38\% F1 in the real deployment environment. To sum up, our main contributions are as follows:
\begin{itemize}[leftmargin=*]
    \item We propose a novel approach (i.e., \tool) to capture the structure information of the commit to accurately learn the semantics for security patch identification. 
    \item We design the contextual AST-based encoder to take the AST paths, which are not limited within the changed code, but further include the paths related to the changed code to capture the contextual information of the changed code.
    \item We conduct an extensive evaluation to demonstrate the effectiveness of \tool on the existing dataset and from the real deployment environment. The experimental results demonstrate that \tool significantly outperforms current state-of-the-art approaches. We have deployed our tool in industry for production.
\end{itemize}

\section{Background}\label{sec:background}
In this section, we briefly introduce the background of commit, abstract syntax tree (AST), and graph neural networks (GNNs) that we will use in this paper.
\subsection{Commit}\label{sec:commit}
A commit, which usually consists of a commit message and a diff, records the modification between different software versions. Due to the various phases of the software development life cycle, a commit can serve for various purposes such as implementation logic, building configurations, and bug fixing. The commit message is used to illustrate the reasons for the modification in natural language, while the diff show the difference between different software versions {(i.e., current vs previous)}, which may involve the modification of multiple files and functions. 
\begin{figure}[tp]
     \centering
     \includegraphics[width=8.8cm]{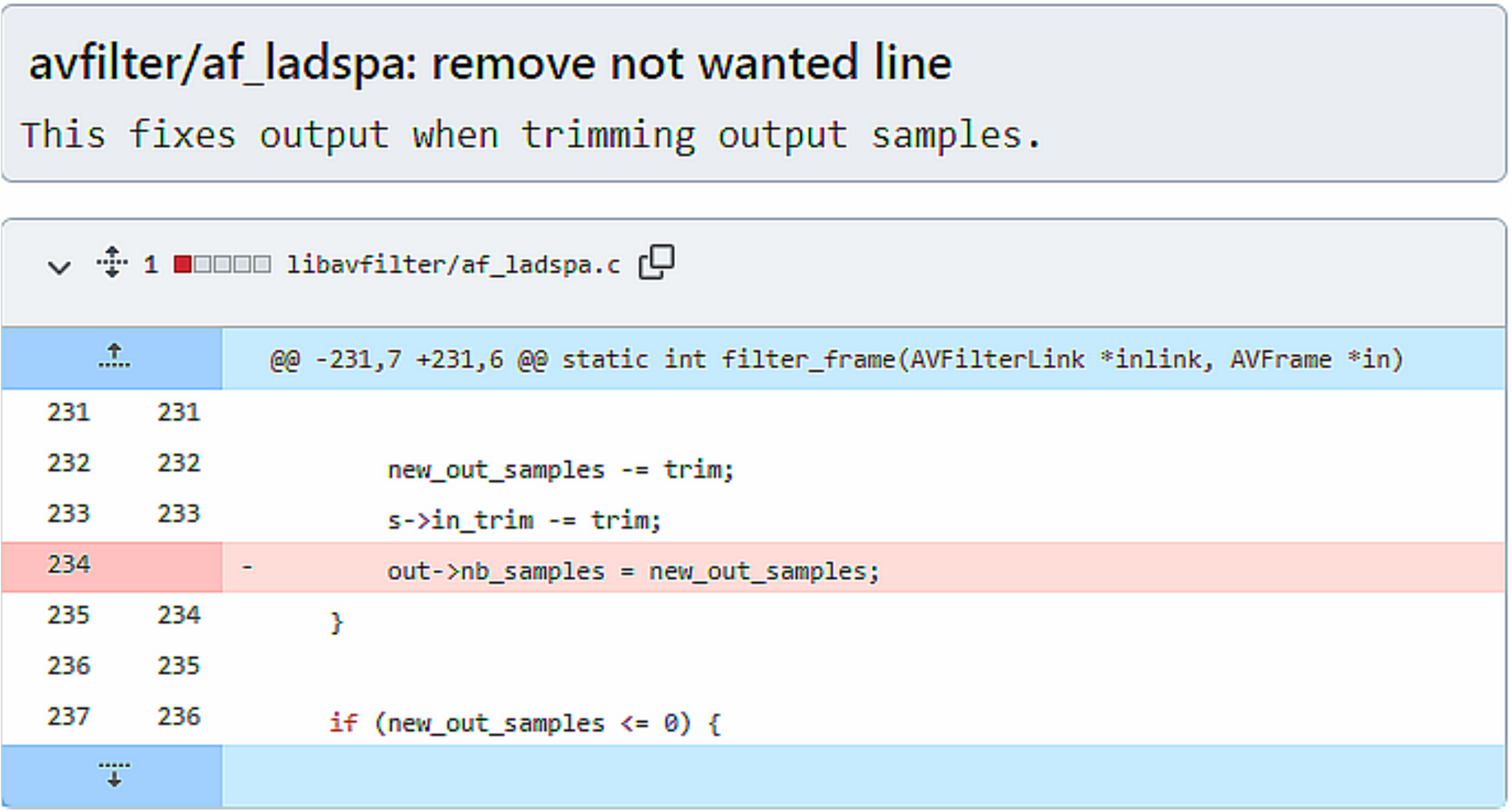}
    \caption{A non-security commit from FFmpeg with its commit id \href{https://github.com/FFmpeg/FFmpeg/commit/ee9345e9054c78ada2e3eda95750c2b4457fc462}{\textit{ee9345e}}.}
     \label{fig:non_commit}
\vspace{-4mm}
\end{figure}
\begin{figure}[tp]
     \centering
     \includegraphics[width=8.8cm]{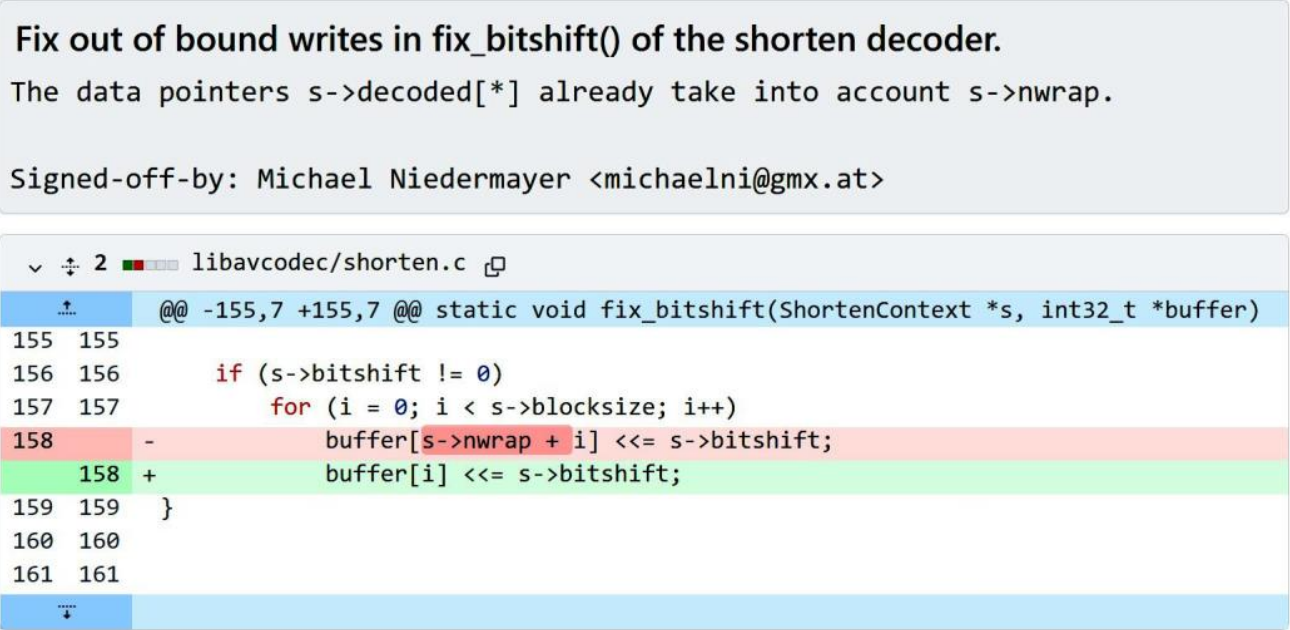}
    \caption{A security commit from FFmpeg with its commit id \href{https://github.com/FFmpeg/FFmpeg/commit/f42b3195d3f2692a4dfc0a8668bb4ac35301f2ed}{\textit{f42b319}}.}
     \label{fig:commit}
\vspace{-4mm}
\end{figure}

\wbz{We present a non-security and security sample commit from the open source project ``FFmpeg'' in Fig.~\ref{fig:non_commit} and Fig.~\ref{fig:commit} respectively to visually demonstrate their difference. To illustrate each component in a commit with more details, we choose the security commit in Fig.~\ref{fig:commit} as an example. }
The content in the upper rectangle is the commit message, which describes the purpose of the current modification in natural language. For this example, the modification serves to fix the bug that an array may be out of the boundary. The lower rectangle is the diffs. We can see that it consists of file name (i.e., ``libavcodec/shorten.c'') and the modified chunk from line 155 to 161 in file ``shorten.c''.
The first line starting with ``@@'' in a chunk is to show some specific information, such as the start modified line number and the method name in the original file.
The code change consists of the content marked with ``+'' in the updated file and its previous version marked with ``-'' in the original file to record the changed code. For example, there is a code change on line 158 in ``libavcodec/shorten.c'', from ``buffer[s\verb|->| nwrap + i] \verb|<<=| s \verb|->| bitshift;'' to ``buffer[i] \verb|<<=| s \verb|->| bitshift;''in Fig.~\ref{fig:commit}. The remaining contents in a chunk are the context about the code changes to reveal the contextual information for the changed code, such as the content from line 156 to line 157. Here, we use a simple commit with only one chunk as a sample for illustration. For other cases, a commit may consist of multiple chunks in one file or multiple changed files.

\begin{figure*}[hbtp]
     \centering
     \includegraphics[scale=0.45]{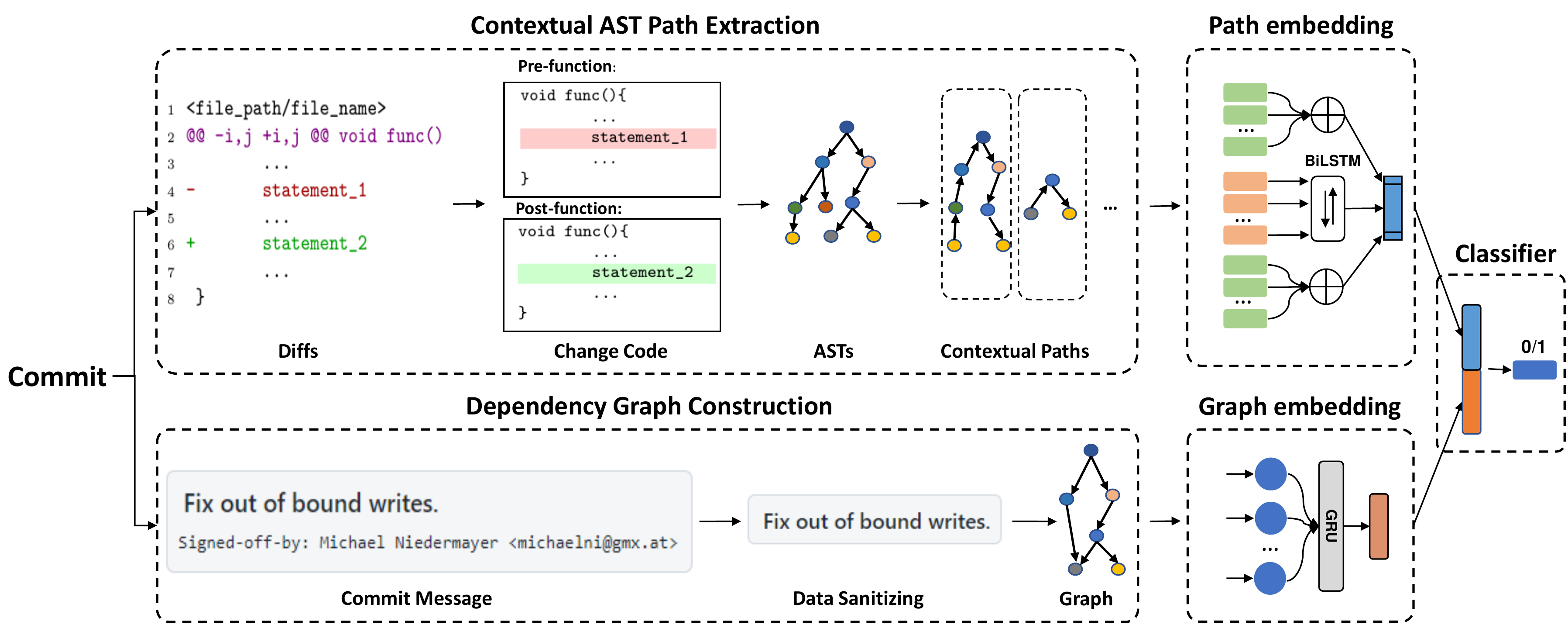}
     \caption{The overview of our framework for security patch identification.}
     \label{fig:overview}
\end{figure*}

\subsection{Abstract Syntax Tree}
Abstract Syntax Tree (AST) is a tree representation of the abstract syntactic structure of source code, while removing unnecessary information such as punctuation and delimiters (e.g., braces, semicolons, parentheses). Hence, AST can be regarded as a high-level abstraction of the source code, which allows program analysis to ignore the grammatical details and focus on the program semantics. A simple AST example is shown in Fig.~\ref{fig:ast}, we can see that it has two types of nodes: terminal nodes and non-terminal nodes, where terminal nodes usually refer to user-defined tokens and represent the names of identifiers from the code such as ``buffer'', ``i''; non-terminal nodes represent a limited set of syntactic structures in a programming language such as ``if\_stmt'', ``func\_declarator''. Since AST is an abstraction of the source code without redundant information, many works~\cite{alon2018code2seq,leclair2020improved,liu2020atom, zhang2019novel, alon2019code2vec} utilize it as the program representation for {various} code-learning tasks.

\subsection{Graph Neural Networks}
Graph Neural Networks (GNNs)~\cite{li2015gated, hamilton2017inductive, kipf2016semi, xu2018graph2seq} can model graph-structured data, which contain nodes and the relations among them, making them popular in different domains such as the social network~\cite{fan2019graph}, molecules~\cite{hamilton2017inductive} and the program~\cite{allamanis2017learning}. GNNs utilize the message passing mechanism to communicate neighborhood information among different type of edges. Based on the different design of message passing, there are a variety of GNN variants such as gated graph recurrent network (GGNN)~\cite{li2015gated}, graph convolutional network (GCN)~\cite{kipf2016semi} and graph attention networks (GAT)~\cite{velivckovic2017graph}. Since a program can be parsed into a structural graph with comprehensive semantics~\cite{yamaguchi2014modeling, allamanis2017learning}, GNNs has achieved promising results on many code-learning tasks such as code summarization~\cite{leclair2020improved, liu2021retrievalaugmented}, code search~\cite{liu2021graphsearchnet} and vulnerability detection~\cite{zhou2019devign}.

\section{Approach}\label{sec:approach}

\begin{figure*}[tp]
    \centering
    \subcaptionbox{Pre-function $f_s$.\label{fig:vuln}}
    {
    \includegraphics[scale=0.8]{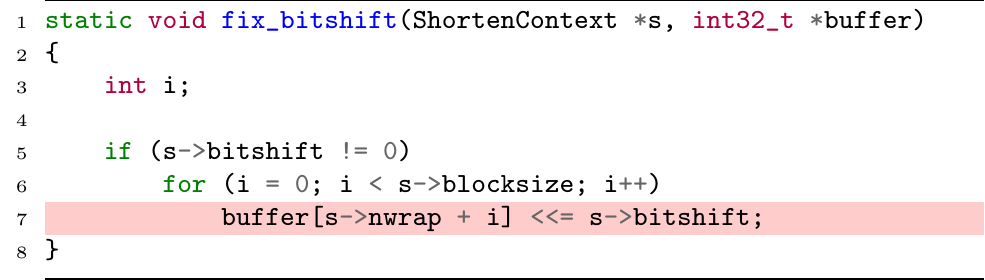}
    }\hspace{8mm}
    \subcaptionbox{Post-function $f_a$.\label{fig:patch}}
    {
    \includegraphics[scale=0.8]{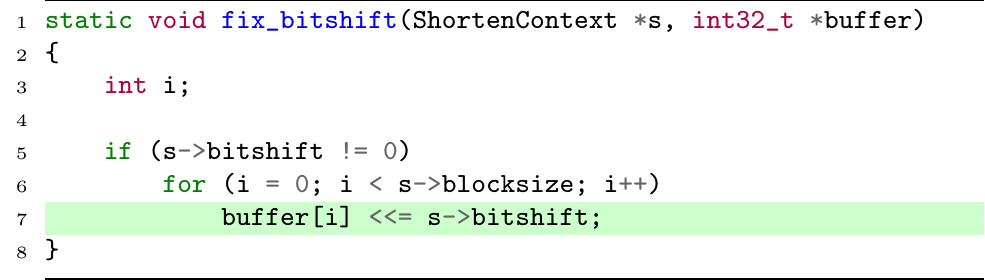}
    }
    \caption{The complete function $f_s$ and $f_a$ extracted from the commit in Fig.~\ref{fig:commit}.}
\label{fig:function_code}
\end{figure*}

In this section, we first formulate the security patch identification problem and then introduce an overview of our approach. We detail each component of our well-designed network for further illustration.

\subsection{Problem Formulation}
In this paper, similar to Zhou et al.~\cite{zhou2021spi}, \tool aims at identifying a commit is a security patch or not. Formally, given a dataset {$D = \{(c,y)|c \in \mathcal{C}, y \in \mathcal{Y}\}$}, where $\mathcal{C}$ is a set of commits, $\mathcal{Y} \in \{0,1\}$ is the set of labels with 1 to indicate that the commit is the security patch and 0 for the non-security patch. \tool formulates it as a binary classification problem and aims to learn a mapping function $f:\mathcal{C} \rightarrow \mathcal{Y}$ for automated identification. The prediction function $f$ can be learned by minimizing the loss function as follows:
\begin{equation}
   \mathrm{min} \sum_i^n \mathcal{L}(f(y_i|c_i))
\end{equation}
where $\mathcal{L(\cdot)}$ is the loss function and we choose the cross entropy to optimize the learned weights, $n$ is the total number of commits in the training set.

\subsection{Overview}
The overview of \tool is shown in Fig.~\ref{fig:overview}, we can see that it jointly utilizes diffs and commit message for the security patch identification. Specifically, to capture the context of the code changes, we first retrieve the corresponding modified functions about the code changes, which are defined as pre-functions and post-functions accordingly. Then, we extract the corresponding contextual AST paths related to the changed code on ASTs, which are parsed from these retrieved functions, to capture the structure information. We feed these extracted AST paths to a code change encoder to learn the representation for the changed code. For the commit message, \tool constructs the dependency graph to capture the token relations in the sequence and uses a commit message encoder to capture the structure information behind the text. Then, \tool concatenates the vector representations produced by the code change encoder and the commit message encoder with a fully connected layer to predict whether the commit is a security patch or not.

\subsection{AST-based Code Change Encoder}
Most of the existing works~\cite{zhou2021spi, wang2021patchrnn} for security patch identification ignore the structure information hidden in the changed code, which limited to achieve the promising results. In addition, they mostly relied on the changed code (e.g., the statement of the line marked with ``+'' or ``-'' in Fig.~\ref{fig:commit}) for embedding, while the contextual information related to the changed code is missed.

To overcome these limitations, 
we propose a well-designed AST-based code change encoder which consists of two sequential parts (i.e., contextual AST path extraction and path embedding).

\subsubsection{Contextual AST Path Extraction}\label{sec:path-extraction}
To capture the contextual information for the changed code, we first retrieve the complete functions {changed by the commit}.
Specifically, we extract the corresponding subtractive statements marked with ``-'' and define them as $d_s$. Similarly, we also extract the additive statements marked with ``+' and define them as $d_a$. Based on $d_s$ and $d_a$, by line number and file name, we can retrieve their original functions defined as the pre-function $f_s$ and the post-function $f_a$. For example, considering the commit in Fig.~\ref{fig:commit}, $d_s$ and $d_a$ are ``buffer[s\verb|->| nwrap + i] \verb|<<=| s \verb|->| bitshift;'' and  ``buffer[i] \verb|<<=| s \verb|->| bitshift;'', respectively, with the given line number (e.g., 158) and file name (e.g., ``bavcodec/shorten.c''), we can successfully retrieve their original functions $f_s$ and $f_a$, which are shown in Fig.~\ref{fig:function_code}. 

\begin{figure}[tp]
    \centering
    \includegraphics[scale=0.5]{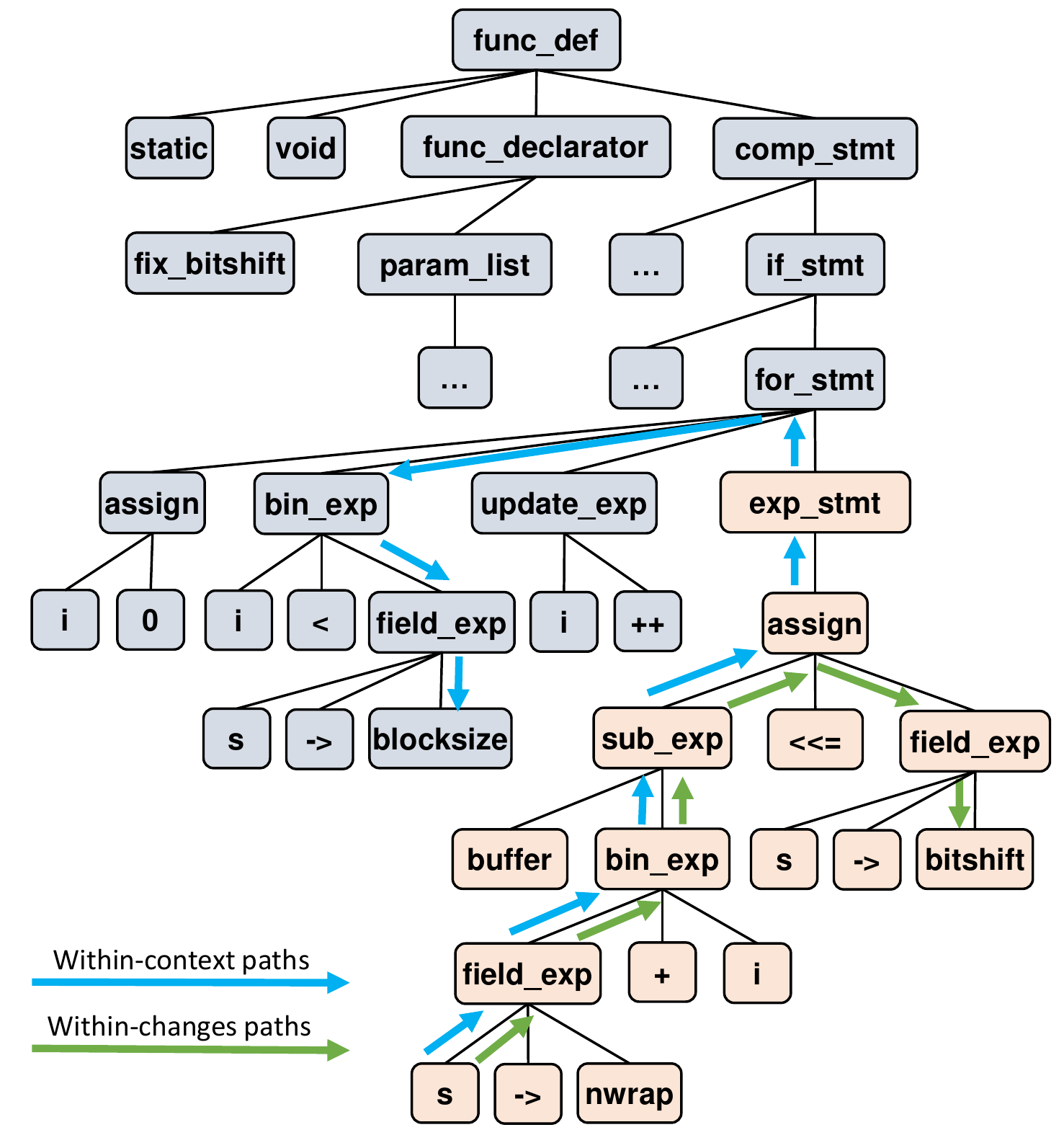}
    \caption{The extracted AST paths from within-changes and within-context, where the function is plot in Fig.~\ref{fig:vuln}.}
    \label{fig:ast}
\end{figure}

Inspired by Code2Vec~\cite{alon2019code2vec} and Code2Seq~\cite{alon2018code2seq}, which parsed the function into AST and constructed the paths between the start node and end node (both of them are the terminal nodes on AST) to capture the structure information behind the function text, we also convert the original functions (i.e., $f_s$ and $f_a$) into ASTs by adopting \textit{tree-sitter} to construct the AST paths. However, unlike their work, which focused on the function-level data, our approach is customized to construct the paths for commits. Specifically, the start node is located at the changed code, while the end node comes from different places based on specific considerations. Furthermore, the end node can be categorized into two groups:
\begin{itemize}[leftmargin=*]
    \item Within-Changes: We include the end node in the changed code to capture the structure information of the changed code. For example, as shown in Fig.~\ref{fig:vuln}, the subtractive statement contains two variables ``s'' and ``bitshift'', hence, there is an AST path\footnote{The shortest path between the start node and the end node.} starts from the ``s'' and ends at the ``bitshift'', which is marked with green arrows in Fig.~\ref{fig:ast}.

    \item Within-Context: To capture the contextual information for the changed code, we randomly select the end nodes, which are from the context for the enhancement. \wbz{Here, we define the contextual information as AST paths where the start node belongs to the changed code and the end node is from other code in a function. } For example, there is a variable ``blocksize'' and it is not in $d_s$, however, it is included in the context of $d_s$, which is shown in Fig.~\ref{fig:vuln}, we also construct a path {between the ``s'' and} ``blocksize'', which is marked with {blue arrows} in Fig.~\ref{fig:ast}.
    
\end{itemize}

\wbz{For any commit $c$, we first extract AST paths of subtractive statements and additive statements, respectively. Then we randomly sample a set of AST paths $\{x_1,\cdots,x_k\}$ from within-changes paths and within-context paths at a ratio of 1:1, where $k$ is their total number. We will discuss the value of $k$ and the ratio in the discussion part of Section \ref{sec:discussion}.}

\subsubsection{Path Embedding}
\begin{figure}[tp]
    \centering
    \includegraphics[scale=0.65]{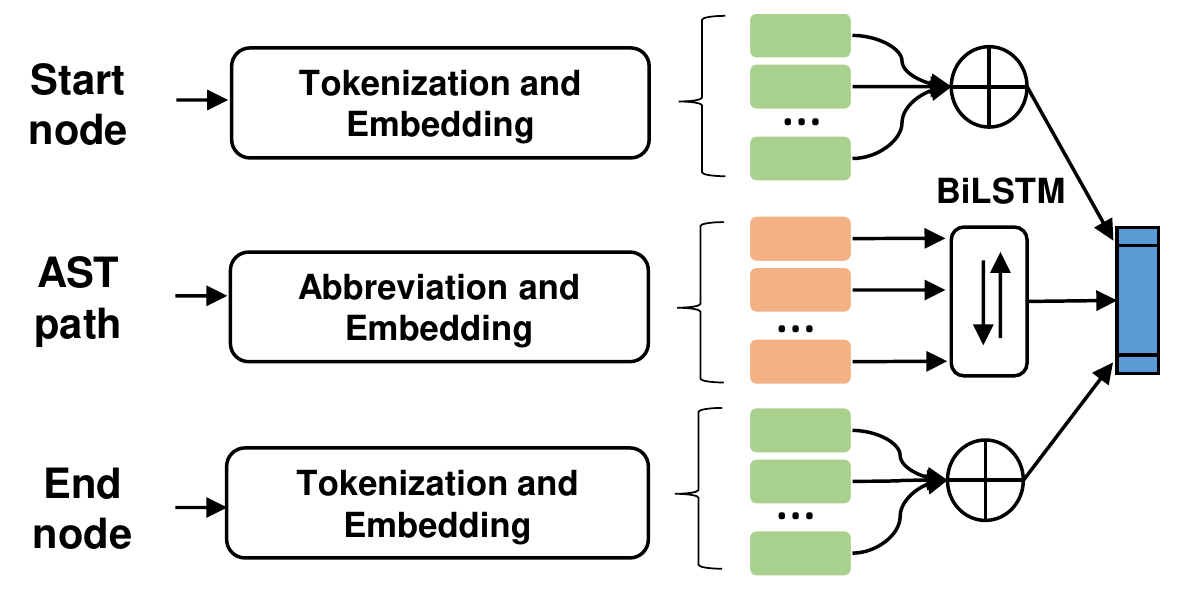}
    \caption{The architecture of the path embedding.}
    \label{fig:code_encoder}
\end{figure}

Given a set of AST paths $\{x_1,\cdots,x_k\}$, each path can {be defined} as $x_i = v_1^iv_2^i\cdots v_l^i$, where $l$ {indicates the total number of nodes}, $v_1^i$ and $v_l^i$ are the start node and end node of the $i$-th path in the AST path set. The start node and the end node are the terminal nodes of AST, which consists of node type and node value, while the non-terminal nodes in the AST path just have the node type. We separately encode the AST path and the terminal nodes as shown in Fig.~\ref{fig:code_encoder}.

\begin{figure*}[tp]
     \centering
     \includegraphics[scale=0.45]{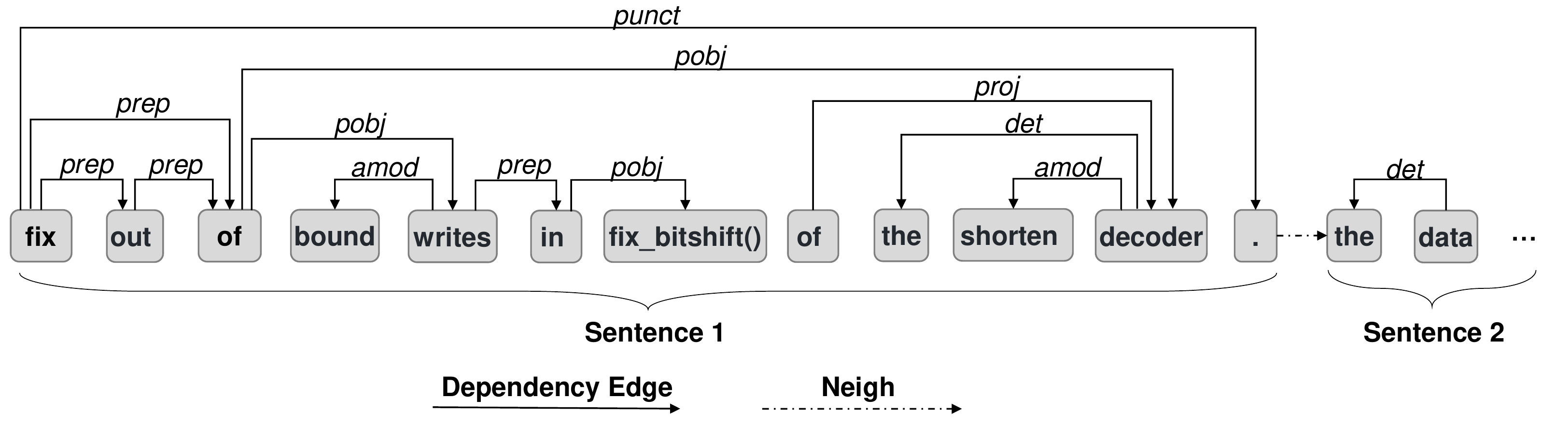}
     \caption{The constructed dependency graph for the commit message in Fig.~\ref{fig:commit}.}
     \label{fig:graph}
\end{figure*}

\begin{itemize}[leftmargin=*]
    \item Path Representation: There are a total number of 220 different symbols produced by \textit{tree-sitter} to represent the node type in AST. However, some of the types may be composed of more than one token, such as ``function\_definition''. {To handle this type of cases, we} simplify them to abbreviations. For example, the node type ``function\_declaration'' is simplified to ``FuncDef''. After that, we use a learnable embedding matrix $\boldsymbol E^{\mathrm{type}} \in \mathbb{R}^{v_1 \times d}$ to encode each node type, where $v_1$ is the length of the vocabulary set for the node type, $d$ is the dimensional length, and feed them into the bidirectional LSTM (BiLSTM) to obtain the final hidden states, which can be formulated as follows:
    \begin{equation}
    \boldsymbol{h}_{v_1^i},\cdots,\boldsymbol{h}_{v_l^i} = \mathrm{BiLSTM}(\boldsymbol E^{\mathrm{type}}_{v_1^i}, \cdots, \boldsymbol E^{\mathrm{type}}_{v_l^i})
    \end{equation}
    \begin{equation}
    \boldsymbol{r}_{path} = [\boldsymbol{h}_{v_l^i}^{\rightarrow};\boldsymbol{h}_{v_1^i}^{\leftarrow}]
    \end{equation}
where $l$ is the total number of nodes in an AST path.
    \item Token Representation: The node $v_1^i$ and $v_l^i$ in $x_i$ are terminal nodes and their values are tokens from the diffs. We split the values into a set of subtokens $S_{subtoken}$ to reduce the vocabulary set. We further represent each subtoken with another learnable embedding matrix $ \boldsymbol E^{\mathrm{subtokens}} \in \mathbb{R}^{v_2 \times d}$, where $v_2$ is the length of the vocabulary set for the subtokens, and then sum them up to represent the terminal {nodes}. For example, for the node $v_1^i$, the calculation can be represented as follows:
    \begin{equation}
        \boldsymbol{r}_{v_1^i} = \sum_{s \in S_{\mathrm{subtoken}} } \boldsymbol E^{\mathrm{subtokens}}_s
    \end{equation}
\end{itemize}

Finally, we concatenate the vector representation of start node (i.e., $\boldsymbol{r}_{v_1^i}$) and end node (i.e., $\boldsymbol{r}_{v_l^i}$) with the AST path (i.e., $\boldsymbol{r}_{path}$), and apply a fully-connected layer followed by a layer normalization to obtain the embedding of the entire AST path.
\begin{equation}
    \boldsymbol{v}_i = \mathrm{Norm}(\mathrm{FC}([\boldsymbol{r}_{v_1^i};\boldsymbol{r}_{path};\boldsymbol{r}_{v_l^i}]))
\end{equation}

Since the code changes are represented by a set of $k$ AST paths, we apply the max-pooling operation over 
{the embedding vector of the AST paths}
i.e., $ \boldsymbol{v}_\mathrm{{c}} = \mathrm{maxpool}(\{\boldsymbol{v}_i\}_{i=1}^k)$ to get the final representation.

\subsection{Graph-based Commit Message Encoder}
Apart from the changed code, a commit also contains a commit message, which illustrates the purpose of the current changes. Compared with {existing approaches} ~\cite{zhou2021spi, wang2021patchrnn}, which {feed} the message into a BiLSTM module to capture the sequential dependency in the sequence, \tool innovates it by constructing the dependency graph with the graph neural network to capture the relations between tokens in the sequence.

\subsubsection{Dependency Graph Construction} 
The dependency parse tree~\cite{de2006generating} describes the dependency relations among words in a sentence by directed linked edges, which has been widely used in word relation extraction~\cite{attardi2014dependency, chen2019reinforcement} in NLP. To better learn the semantics in the commit message, we {construct the dependency graph with two steps}: data {sanitization} and graph construction. For data {sanitization}, since software developers may record some information not related to the code changes, such as website links, signatures, {or} emails. To sanitize the commit message, we remove these {irrelevant} information in commit message by the regular expression. Then, we construct the dependency graph according to the cleaned commit message. 

Specifically, {given the cleaned commit message $m$ consists of multiple sentences}, for each sentence in $m$, we utilize \textit{Spacy} to construct its dependency tree. We further connect the neighboring dependency parse trees by connecting those nodes that are at the end and the beginning of two adjacent sentences with the edge ``neigh'' and construct the dependency graph $\mathcal G = (\mathcal V, \mathcal E)$, where $\mathcal V$ is a set of nodes in the graph and $\mathcal E$ denotes the relations (i.e., edges) within the nodes. Furthermore, each node in the graph is also the original token in $m$ and there are 49 different edge types to describe these token relations. We take the commit message in  Fig.~\ref{fig:commit} as an example and the constructed dependency graph is shown in Fig.~\ref{fig:graph}. We can see that a token may have different semantic relations linked to other tokens. For example, there is an edge ``prep'' points from ``fix'' to ``out'' and ``of'' to describe the prepositional modifier of the verb ``fix''. Furthermore, to connect two different sentences, we further add an edge ``neigh''. For example, as shown in Fig.~\ref{fig:commit}, the edge ``neigh'' connects the nodes that are the last token ``.'' in the first sentence and the first token ``the'' in the second sentence.

\begin{figure}[tp]
    \centering
    \includegraphics[scale=0.65]{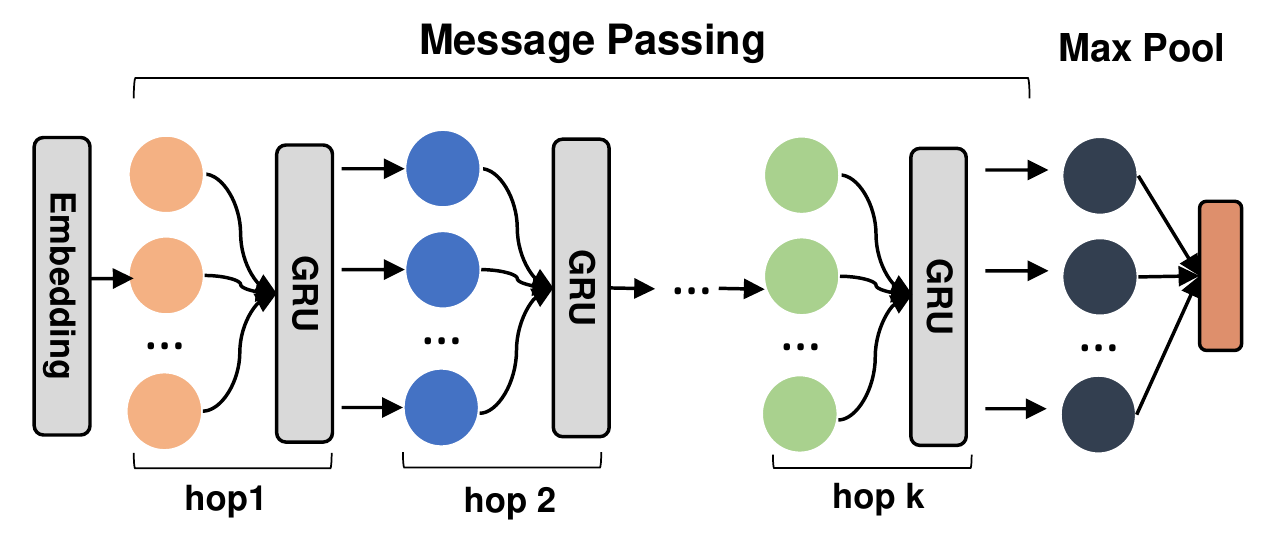}
    \caption{The architecture of the graph embedding.}
    \label{fig:message_encoder}
\end{figure}

\subsubsection{Graph Embedding}
Given the dependency graph $\mathcal G = (\mathcal V, \mathcal E)$ to describe the relations in the tokens, we need to learn its representation. A variety of graph neural networks (GNNs)~\cite{li2015gated, velivckovic2017graph, kipf2016semi} can be chose, inspired by the recently advanced Gated Graph Neural Network on modelling the program~\cite{allamanis2017learning, allamanis2020typilus, liu2021retrievalaugmented, liu2021graphsearchnet, fernandes2018structured}, we also select GGNN to learn the vector representation, as shown in Fig.~\ref{fig:message_encoder}.
Specifically, each node $v \in \mathcal V$ is initialized by a learnable embedding matrix $\boldsymbol E \in \mathbb{R}^{v_3 \times d}$, where $v_3$ is the length of the vocabulary set for the commit message and $d$ is the dimensional length, and gets its initial vector $\boldsymbol{h}^0_v \in \mathbb{R}^d$. We apply a fix number of hops (i.e., $T$) to propagate the node information along the edges. At each computation hop $t$, where $ 1 \leq t \leq T$, we select the summation function as the aggregation function to aggregate the neighboring node features from the previous hop, which can be expressed as follows: 
\begin{equation}
    \boldsymbol{h}_{N(v)}^t = \mathrm{SUM}(\{\boldsymbol{h}^{t-1}_u|\forall u \in N_{(v)}\})
\end{equation}
where $N_{(v)}$ is a set of neighborhood nodes that are connected with $v$. Then, we use a Gated Recurrent Unit (GRU)~\cite{cho2014learning} to update the node representation as follows:
\begin{equation}
    \boldsymbol{h}_v^t = \mathrm{GRU}(\boldsymbol{h}^{t-1}_v,\boldsymbol{h}^t_{N_{(v)}})
\end{equation}

After a fixed number of hops (i.e., $T$), we obtain the final node representation $\boldsymbol{h}_v^T$ and apply the max-pooling over all nodes $\{\boldsymbol{h}_v^T |\forall v \in V\}$ to obtain the graph representation $\boldsymbol{v}_m$, defined as follows:
\begin{equation}
    \boldsymbol{v}_m = \mathrm{maxpool}(\mathrm{FC}(\{\boldsymbol{h}^{T}_v |\forall v \in V\}))
\end{equation}
where $\mathrm{FC}$ is a fully connected layer. The graph representation $\boldsymbol{v}_m$ can be considered as the vector representation for the cleaned commit message $m$.

\subsection{Ensembling}
By the code change encoder and commit message encoder, we can get the vector representation for the changed code (i.e., $\boldsymbol{v}_\mathrm{c}$) and commit message (i.e., $\boldsymbol{v}_m$) respectively. We further ensemble both to represent the commit $c$. Specifically, we concatenate the graph embedding $\boldsymbol{v}_\mathrm{m}$ with the code change embedding $\boldsymbol{v}_\mathrm{c}$, and apply a fully connected layer to the classifier:
\begin{align}
    \mathrm{prob} = \mathrm{Sigmoid}(\mathrm{FC}([\boldsymbol{v}_m;\boldsymbol{v}_\mathrm{c}])) \\
    \mathrm{prediction} =
\left\{
	\begin{array}{ll}
		1 & \mbox{if } \mathrm{prob} \geq 0.5 \\
		0 & \mbox{if } \mathrm{prob} < 0.5
	\end{array}
\right.
\end{align}
where $\mathrm{prob}$ denotes the probability obtained by the activation function $\mathrm{Sigmoid}$, $\mathrm{prediction}$ is the predicted result where 0 for $\mathrm{prob}$ less than 0.5 and 1 for the others.

\section{Evaluation Setup}\label{sec:setup}
In this section, we first introduce the dataset used, followed by the selected baselines for the comparison, then present the evaluation metrics with the model settings of \tool.
\subsection{Dataset}\label{sec:dataset}
Currently, to the best of our knowledge, in addition to SPI~\cite{zhou2021spi} making their data partially public (i.e., FFmpeg and QEMU) are available, the other works for security patch identification have not released the data so far. 

\wbz{Hence, to facilitate the evaluation, we borrow the SPI dataset, which is collected from 4 high-profile open source projects (i.e., Linux, FFmpeg, Wireshark, QEMU) on Github with the help of manual analysis and labeling by security analysts. After filtering out those samples that cannot be parsed into ASTs or graphs, we finally get the dataset for evaluation. As shown in Table~\ref{tb:dataset}, the dataset consists of 38,593 labeled commits, where 19,896 of them are security related patches and 18,697 are non-security. } 
Furthermore, the commits of Linux are collected from 2016 to 2017, while the other three projects are up to January 2018. Different {from} SPI, we split the data of each project with the ratio of {80:10:10} and then fuse the divided data of each project as {train/validation/test sets} for evaluation.

\begin{table}[tb]
\centering
\caption{The statistics of dataset}
\scalebox{1.15}{
\begin{tabular}{l|cc|r}
\hline
Project & Security & Non-security & Total \\ \hline
Linux     & 7,358  & 3,144  & 10,502 \\
FFmpeg    & 5,393  & 6,210  & 11,603 \\
QEMU      & 4,256  & 5,516  & 9,772  \\
Wireshark & 2,889  & 3,827  & 6,716  \\\hline
Total     & 19,896 & 18,697 & 38,593 \\\hline
\end{tabular}
}
\label{tb:dataset}
\end{table}

\subsection{Baselines}
To evaluate our approach, we compare \tool with {six} state-of-the-art security patch identification approaches. The details of these {six} approaches are shown as follows:

\noindent \textbf{Stacking}~\cite{zhou2017automated}. Zhou et al. proposed to identify security issues from commit messages and bug reports by learning a stack of six basic classifiers such as SVM, random forest for automated identification. In \tool, since the bug reports are not available, we ignore them and just use the commit messages for the experiments. 

\noindent \textbf{SPI}~\cite{zhou2021spi}. SPI considered both commit messages and code changes as the input. It proposed two separate encoders to learn the representation of commit message and code change respectively and each of them consists of a BiLSTM layer and a CNN layer. Then, two encoders are combined together to get the final probability of predicting whether a commit is vulnerable or not. 

\noindent \textbf{PatchRNN}~\cite{wang2021patchrnn}. Similar to SPI, PatchRNN also considered two separate BiLSTM encoders to detect security patches. In addition to commit message, it takes the entire function code before and after the patch into consideration, rather than the code changes used in SPI. 

\noindent \textbf{Commit2vec}~\cite{lozoya2021commit2vec}. It ignored the commit message and just used the code changes to identify the security-relevant commit, Furthermore, it extracted AST paths, which are related to the code changes as the input to represent the code changes, and fed the paths into the fully connected layers to learn the vector representation for identification.

\noindent \textbf{Transformer}~\cite{vaswani2017attention}. Transformer has achieved great success in different fields. We also include it as a baseline. Specifically, we utilize two separate transformer encoders to encode the commit message and code changes respectively. Each encoder has 6 identical layers, each of them has 8 heads to learn different subspace features. 

\noindent \wbz{\textbf{VulFixMiner}~\cite{zhou2021finding}. VulFixMiner applied CodeBert to extract semantic meaning from code changes to identify silent vulnerability fixes. Similar to commit2vec, VulFixMiner did not consider commit message. }

Except for SPI, which we reproduce the experiments with the official code, none of the other baselines are open source, and we re-implement these baselines following to the original papers strictly and carefully.

\subsection{Evaluation Metrics}
Following SPI~\cite{zhou2021spi}, we use \textbf{precision} (Pre), \textbf{recall} (Rec) and \textbf{F1} as our evaluation metrics, {and} further add \textbf{accuracy} (Acc), since it is a common metric to evaluate classification systems. Before introducing the used metrics, we first {illustrate} four important terms that will be used in computing the metric values:
\begin{itemize}[leftmargin=*]
    \item true\_positives: the cases in which predicted as security patches actually are security patches.
    \item true\_negatives: the cases in which predicted as non-security patches actually are non-security patches.
    \item false\_positives: the cases in which predicted as security patches actually are non-security patches.
    \item false\_negatives: the cases in which predicted as non-security patches actually are security patches.
\end{itemize}

\textbf{Precision} represents the proportion of true positives to the predicted total positives, which infers how precise the model is in identifying security patches. The formula of precision is shown below.
\begin{equation}
   \mathrm{precision} = \frac{\mathrm{true\_positives}}{\mathrm{true\_positives} + \mathrm{false\_positives}} 
\end{equation}

\textbf{Recall} is the ratio of true positives to all positives in ground truth, which can be seen as a measure of how robust the model is in identifying security patches. It can be calculated as follows.
\begin{equation}
    \mathrm{recall} = \frac{\mathrm{true\_positives}}{\mathrm{true\_positives} + \mathrm{false\_negatives}} 
\end{equation}

\textbf{F1} is the harmonic mean between precision and recall. A high F1 implies low false positive as well as low false negative. In other words, a high F1 means that the model is highly precise and robust. F1 can be computed as follows.
\begin{equation}
    \mathrm{F1} = 2 * \frac{\mathrm{precision} * \mathrm{recall}}{\mathrm{precision} + \mathrm{recall}} 
\end{equation}

\textbf{Accuracy} is the ratio of the number of correct predictions to the total number of input samples. Therefore, a high accuracy implies that the model performs well in identifying both security patches and non-security patches. Accuracy can be formulated as follows.
\begin{equation}
    \mathrm{accuracy} = \frac{\mathrm{true\_positives} + \mathrm{true\_negatives}}{\mathrm{total\_sample}} 
\end{equation}
where total\_sample is the sum of true positives, true negatives, false positives and false negatives.

\subsection{Model Settings}\label{sec:settings}

For the code change encoder, we set the embedding size for the node type and node value to 128. The dimension of hidden states in the BiLSTM is set to 128. The maximum number of AST paths $k$ is set to 500 to represent the code changes. For the commit message encoder, the embedding size of tokens is also set to 128, following the code change encoder. Furthermore, the dimension size of the hidden states in GRU is 128. We use the Adam optimizer with a learning rate of 0.001 to train the model. The early stop is set to 10 which means that the training process is stopped when no improvements on Acc for 10 epochs. 
All experiments are conducted on Intel(R) Xeon(R) Gold server with three Nvidia Graphics Tesla V100. To avoid the effect of fluctuations on the random seed, we repeat the experiments three times with different random seeds and report the average values for all experiments. 
\section{Evaluation Results}\label{sec:results}
To evaluate the effectiveness of \tool, we first compare the performance of \tool against the state-of-the-art approaches and then conduct a deep analysis on the effect of contextual AST paths and graph-based commit message encoder. To confirm the validity of \tool, we further investigate its performance in a real deployment environment \wbz{and provide a qualitative analysis of the prediction results.}. Our experiments mainly focus on the following {five} research questions:

\begin{itemize}[leftmargin=*]
    \item RQ1: Can \tool outperform the baselines in the evaluation metrics?
    \item RQ2: Does the contextual AST paths improve the performance in code change encoder?
    \item RQ3: Is the graph-based commit message encoder powerful in learning the commit message?
    \item RQ4: What is the performance of \tool in a real deployment environment?
    \wbz{\item RQ5: The qualitative analysis of the results provided by \tool.}
\end{itemize}

\begin{table}[tp]
\centering
\caption{Evaluation results in percentage compared with the baselines.}
\scalebox{1.15}{
\begin{tabular}{c|cccc}
\hline
Approach    & Acc & Pre & Rec & F1 \\ \hline
Stacking    & 69.35 & 71.50 & 67.44 & 69.41 \\
Commit2vec  & 62.29 & 62.75 & 66.26 & 64.46 \\
PatchRNN    & 81.68 & 83.05 & 81.01 & 82.01 \\
SPI         & 84.95 & 85.82 & 84.82 & 85.32 \\
Transformer & 74.27 & 78.37 & 69.20 & 73.50 \\
VulFixMiner & 63.61 & 63.25 & 65.23 & 64.22 \\\hline
\tool     & \textbf{88.96} & \textbf{87.88} & \textbf{91.26} & \textbf{89.54} \\ \hline
\end{tabular}
}
\label{tb:overall}
\end{table}

\subsection{RQ1:Comparison with the State-of-the-art Approaches}\label{lb:exp1}
We compare \tool with {six} state-of-the-art approaches and the evaluation results are shown in Table~\ref{tb:overall}.

An interesting finding is that the machine learning-based approach Stacking significantly outperforms the deep learning-based approach Commit2vec 
\wbz{and VulFixMiner}, which demonstrates that commit messages play a more critical role than code changes for the security patch identification and performance by just using a stack of multiple machine learning classifiers on the messages can outperform the performance of deep neural networks on the code changes. 
\wbz{After analysis, we find that when developers fix a bug and submit the corresponding patch, they would describe their purpose in detail in commit message. For security patches, their commit message often contains some security-related words, such as "fix buffer overflow" and "fix memory leak", which contain explicit security-related information and are easily recognized by the model. In contrast to commit messages, security information in code changes is implicit and cannot be easily extracted from the source code. Therefore, having commit messages as input can bring more benefits than code changes.}

Furthermore, we find that the performance of Transformer is lower than that of BiLSTM-based models (i.e., PatchRNN and SPI), which is in conflict with the current consensus that transformer has more expression capacity than BiLSTM. We conjecture that it is caused by the limited amount of data used for training. Specifically, the total number of samples in the dataset is only 40,523, which makes transformer overfit to the training data. The smaller size of the dataset impacts the transformer to obtain the promising results compared with the BiLSTM-based models. 

The last row in Table~\ref{tb:overall} shows the results of \tool. We can observe that \tool outperforms the baselines significantly, achieving 88.61\% in accuracy and 89.25\% in F1, exceeding the best baseline PatchRNN by 5.99\% in accuracy and by 5.81\% in F1. The experimental results demonstrate the effectiveness of our designed AST-based code change encoder and graph-based commit message encoder for security patch identification.

\begin{tcolorbox}[breakable,width=\linewidth,boxrule=0pt,top=1pt, bottom=1pt, left=1pt,right=1pt, colback=gray!20,colframe=gray!20]
\textbf{Answer to RQ1:} \tool outperforms {six} state-of-the-art baselines significantly in accuracy and F1, we attribute the improvements to the well-designed code change encoder and the commit message encoder. 
\end{tcolorbox}

\subsection{RQ2:Comparison with Different Code Change Encoders}\label{lb:exp2}
We further conduct experiments to compare the performance of different code change encoders to confirm the effectiveness of our designed AST-based code change encoder. Specifically, we compare with PatchRNN, SPI, Transformer, VulFixMiner and Commit2vec,
where PatchRNN took the entire function code as input with BiLSTM for learning, SPI fed the code changes with a BiLSTM followed by the CNN layer to learn the representation, transformer fed the changed code into the transformer encoder for learning, \wbz{VulFixMiner fed the code changes into a CodeBert to get the representation,} and Commit2vec extracted the AST paths related to the code changes and fed these paths into fully-connected layers for learning.
For these approaches, if they involve the commit message encoder, we turn it off and just use the code change encoder to compare the results. 
In addition, we also ablate the effect of the AST paths that only extracted from the code changes (i.e., \tool w/o Context) and the paths where the end node is from the context (i.e., \tool w/o Changes). The experimental results are presented in Table~\ref{tb:exp2}.

We can observe that our AST-based code changes encoder outperforms the baseline code changes encoders by a significant margin in terms of Acc and F1. Specifically, PatchRNN performs the worst among these approaches, we infer that it is caused by PatchRNN takes the entire function rather than the changed code as the input, which introduces too many noises for patch identification. Considering purely using the changed code for learning, transformer has a better performance than SPI. However, when the commit message encoder is turned on, the performance is lower than SPI in Table~\ref{tb:overall}. We believe it is reasonable because the difficulty of learning the natural language commit message is much lower than the code, which makes the equipped powerful transformer overfit to the commit message encoder. Thus, when combining both encoders, it performs worse than BiLSTM. However, when purely using the transformer for code learning, it has a higher performance than BiLSTM due to the more powerful expression ability of the transformer. \wbz{VulFixMiner performs close to Transformer, as it also leverages a powerful model for patch identification.} However, the performance of the transformer \wbz{and VulFixMiner} is still lower than \tool, we attribute to the structure information used in the changed code of \tool. Furthermore, although Commit2vec has higher accuracy and F1 than SPI, due to that it also encodes AST as input, its performance is lower than \tool. We believe that it is caused by the simple fully connected layer used in Commit2vec that has limited learning capacity compared to BiLSTM used in \tool. 

Lastly, we ablate the effect of the contextual information used in AST paths and the experimental results are shown at the last row of Table~\ref{tb:exp2}. We can see that the paths extracted from the changed code and the context are both beneficial in improving the performance, and when incorporating both, \tool could achieve the best performance.

\begin{tcolorbox}[breakable,width=\linewidth,boxrule=0pt,top=1pt, bottom=1pt, left=1pt,right=1pt, colback=gray!20,colframe=gray!20]
\textbf{Answer to RQ2:} AST-based code change encoder achieves higher performance as compared to the baseline encoders, we attribute to the structure information used in the changed code. Furthermore, we also confirm that the contextual AST paths are also beneficial in improving the performance.
\end{tcolorbox}

\begin{table}[tp]
\centering
\caption{Evaluation results on different code change encoders.}
\scalebox{1.05}{
\begin{tabular}{c|cccc}
\hline
Approach     & Acc  & Pre   & Rec   & F1             \\ \hline
PatchRNN    & 59.07 & 61.43 & 55.38 & 58.25 \\
SPI         & 61.37 & 61.86 & 65.34 & 63.56 \\
Transformer & 62.71 & 60.84 & 77.65 & 68.23 \\
Commit2vec  & 62.29 & 62.75 & 66.26 & 64.46 \\
VulFixMiner & 63.61 & 63.25 & 65.23 & 64.22  \\\hline
\tool w/o Changes  & 64.94 & \textbf{65.14} & 69.61 & 67.30     \\
\tool w/o Context & {63.49} & 64.02 & {66.81} & {65.39} \\ 
\tool & \textbf{65.56} & 63.59 & \textbf{78.27} & \textbf{70.17} \\\hline
\end{tabular}
}
\label{tb:exp2}
\end{table}

\subsection{RQ3:Comparison with Different Commit Message Encoders}
Similar to Section~\ref{lb:exp2}, we also make a comparison with {four} state-of-the-art commit message encoders, including Stacking, PatchRNN, SPI, and Transforme, where Stacking used a stack of 6 basic machine learning classifiers on the extracted features of commit messages, PatchRNN encoded the commit message with a BiLSTM to learn the representation, SPI further added a CNN layer followed by BiLSTM for classification, and Transformer fed the message to the transformer encoder for learning. The experimental results are shown in Table~\ref{tb:exp3}.

We can see that Stacking has the worst performance. We attribute to the limitation by the used {techniques} (i.e., the adopted machine learning techniques) are less advanced than deep neural networks for the same input. Furthermore, transformer has a lower performance than PatchRNN and SPI, which confirms that transformer is easier to overfit to the commit message and thus has a lower performance. Due to the lower performance of the transformer compared to PatchRNN and SPI in the commit message, it further affects its performance when combined with the code change encoder (see Table~\ref{tb:overall}).
Finally, we can find that \tool achieves the highest accuracy, precision, recall and F1 at 87.26\%, 85.04\%, 91.66\% and 88.23\%, which outperforms other approaches significantly. 
It indicates that by constructing the dependency graph to capture the structure information behind the commit message and further utilizing GNNs to learn the token relations, our designed commit message encoder could learn the semantics well, and thus produces promising results. 

\begin{tcolorbox}[breakable,width=\linewidth,boxrule=0pt,top=1pt, bottom=1pt, left=1pt,right=1pt, colback=gray!20,colframe=gray!20]
\textbf{Answer to RQ3:} Graph-based commit message encoder is powerful to produce the promising results, we attribute to the constructed dependency graph with GNNs to learn the structure information behind the text to capture the commit message semantics. 
\end{tcolorbox}

\begin{table}[tp]
\centering
\caption{Evaluation results on different commit message encoders.}
\scalebox{1.15}{
\begin{tabular}{c|cccc}
\hline
Approach     & Acc            & Pre            & Rec            & F1             \\ \hline
Stacking    & 69.35 & 71.50 & 67.44 & 69.41 \\
PatchRNN    & 81.84 & 84.41 & 79.45 & 81.85 \\
SPI         & 83.92 & 84.98 & 83.58 & 84.27 \\
Transformer & 79.43 & 82.86 & 75.78 & 79.16 \\\hline
\tool     & \textbf{87.66} & 85.53 & \textbf{91.66} & \textbf{88.49} \\ \hline
\end{tabular}
}
\label{tb:exp3}
\end{table}

\begin{table}[tp]
\centering
\caption{Evaluation results on unseen commits from the deployment.}
\scalebox{1.15}{
\begin{tabular}{c|cccc}
\hline
Approach    & Acc            & Pre            & Rec            & F1             \\ \hline
Stacking    & 58.45          & 65.81          & 51.35          & 57.69          \\
Commit2vec  & 55.65          & 59.42          & 54.53          & 56.87          \\
PatchRNN    & 81.57          & \textbf{90.53} & 74.36          & 79.05          \\
SPI         & 81.87          & 91.51          & 73.99          & 81.82          \\
Transformer & 67.27          & 63.61          & \textbf{95.01} & 76.20          \\ 
VulFixMiner & 56.13          & 58.65          & 56.03          & 57.31          \\ \hline
\tool     & \textbf{86.81} & 84.49          & 91.52          & \textbf{87.86} \\ \hline
\end{tabular}
}
\label{tb:exp4-1}
\end{table}

\subsection{RQ4:Evaluation in a Real Deployment Environment}\label{sec:wild}
We discuss the efficiency and practicality of \tool in a real production setting. We deploy \tool with the baselines on an industrial platform from our industry collaborator and investigate the real performance in terms of detection accuracy and efficiency. Specifically, we deployed the pipeline of each approach to the real-time platform, and {each pipeline consists of the scripts of data pre-processing for commits, the pre-trained model with scripts for prediction.}

We first evaluate the accuracy among different approaches on the unseen commits labeled by professional security experts from industry. As shown in Section~\ref{sec:dataset}, the dataset used consists just of commits before January 2018. {Hence, we further collect the commits from February 2018 to March 2021 from the four open source projects (that is, Linux, FFmpeg, QEMU, and Wireshark) with the help of four security experts.} Following SPI~\cite{zhou2021spi}, the keyword filtering process and cross-validation by experts are carried out to filter irrelevant commits. After the strict validation, there are 6,297 samples in total, of which 3,517 are the security related patches and 2,780 are non-security for the four projects. We test the performance of different approaches on these data and show the experimental results in Table~\ref{tb:exp4-1}. We can observe that all approaches suffer from a decline over the evaluation metrics, compared with the results in Section~\ref{lb:exp1}. It is reasonable, since the diversity of the newly collected samples is increased in the real deployment environment. In contrast, for a specific dataset, where the data distribution over this dataset is fixed, the model tends to perform better compared to deploying it to the unseen data. However, although performance is decreased on these approaches, \tool could still achieve better performance in terms of accuracy and F1 compared to other baselines, which confirms the effectiveness of our approach.

To evaluate detection efficiency, we average the prediction time, which includes the data pre-processing time and inference time, for the collected commits, and the time spent for predicting one commit by each approach is presented in Table~\ref{tb:exp5}, where the ``Extraction'' column defines the time for the commit-related function extraction, ``Processing'' column defines the processing time to convert the data into an appropriate format for the model, ``Inference'' column defines the time for model inference and ``Total'' column is the sum time for the previous three columns. Because the approach Stacking, SPI, Transformer and VulFixMiner only utilize changed code as the input, the time for the function extraction is zero. From Table~\ref{tb:exp5}, we can see that \tool requires longer time to make a prediction, which indicates the {higher} complexity of our approach. Furthermore, we can find that the pre-processing occupies most of the time in our approach, since it is relatively complicated and involves many operations such as AST path extraction and message dependency graph construction. However, generally, we can {still} conclude that the time for a prediction by our approach is acceptable {since} it only takes around 0.2 second to produce a result.

\begin{tcolorbox}[breakable,width=\linewidth,boxrule=0pt,top=1pt, bottom=1pt, left=1pt,right=1pt, colback=gray!20,colframe=gray!20]
\textbf{Answer to RQ4:} Although the accuracy of \tool decreases in a real deployment environment where the data distribution is different from the used dataset, the performance of \tool is still higher than the baselines. Furthermore, the efficiency of our approach (i.e., the prediction time) is also acceptable compared to the baselines.
\end{tcolorbox}

\begin{table}[tp]
\centering
\caption{Time spent in millisecond (ms) by each approach. }
\scalebox{1}{
\begin{tabular}{c|ccc|c}
\hline
Approach           & Extraction     & Processing                   & Inference    & Total           \\ \hline
Stacking         & 0.00           & 0.30                         & \textbf{7.43} & 7.73            \\
Commit2vec       & 10.75          & 195.33                       & 2.53          & 208.61          \\
PatchRNN         & 10.75 & 0.73                         & 1.18          & 12.66           \\
SPI              & 0.00          & 0.57 & 1.52          & 2.09            \\
Transformer      & 0.00           & 0.57 & 1.71          & 2.28            \\ 
VulFixMiner      & 0.00           & 0.57 & 1.66          & 2.23            \\ \hline
\tool         & \textbf{10.75} & \textbf{199.41}              & 4.07          & \textbf{214.23} \\ \hline
\end{tabular}
}
\label{tb:exp5}
\end{table}

\subsection{RQ5: A Qualitative Analysis of Results by \tool.}\label{sec:wild}
\wbz{From the previous experimental results, we can conclude that \tool performs well in accuracy and F1. Furthermore, we supplement a qualitative analysis of the results obtained by \tool to understand the reason for the correct predictions and misclassifications. Since the prediction results by \tool include 1,761 true positive, 1,688 true negative, 242 false positive and 169 false negative samples, we randomly select 100 samples from each class for manual analysis. After the analysis, we discover the following phenomena:}
\begin{itemize}[leftmargin=*]
    \item \wbz{Compared to those true negative samples (i.e., correctly identified as non-security patches), most of the true positive samples (i.e., correctly identified as security patches) contain security-related words (i.e., vulnerability noun, such as ``buffer overflow'') in the commit message or have security patterns in the change code. Specifically, 75 out of 100 true positive samples contain obvious security-related words in commit message, such as ``fix memory leak'', ``avoid integer overflow'' and ``fix division by zero''. For example, the commit \href{https://github.com/FFmpeg/FFmpeg/commit/ff763351e74550df3b9a0465634d1ec48b15b043}{\textit{ff76335}} with commit message ``avfilter/zscale: fix memory leak.'' and the commit \href{https://github.com/qemu/qemu/commit/51c1ebb1bc2642296379a8db1ba9dfb4f78a2f80}{\textit{51c1ebb}} with commit message ``Fix SCSI off-by-one device size.''. In addition, 20 true positive samples do not contain obvious security-related words in the commit message, but have security patterns in the change code, such as adding condition check and modifying condition check. For example, the commit  \href{https://github.com/FFmpeg/FFmpeg/commit/066dc0437368acd21546ee327a0a94c60c3808b2}{\textit{066dc04}} adds condition check to avoid null pointers, as shown in Fig.~\ref{fig:rq5-1}.}
    
    \begin{figure}[t]
    \centering
    \includegraphics[width=8cm]{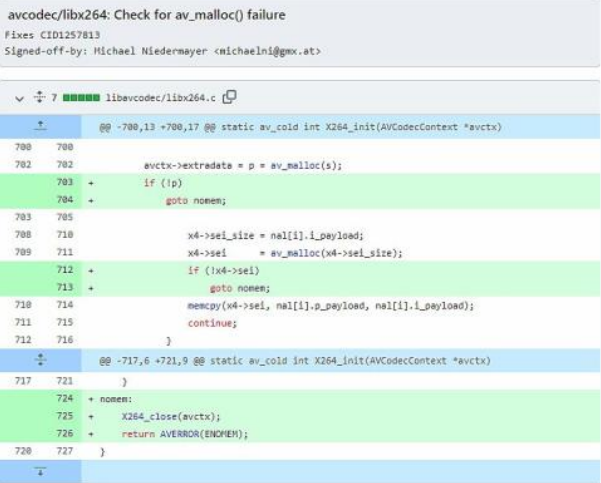}
    \caption{Commit \href{https://github.com/FFmpeg/FFmpeg/commit/066dc0437368acd21546ee327a0a94c60c3808b2}{\textit{066dc04}}. }
    \label{fig:rq5-1}
    \vspace{-4mm}
    \end{figure}

    \item \wbz{For those misclassified samples, we find that they may also be related to security-related words in the commit message. Specifically, some of the false positive samples (i.e., misclassified as security patches) contains security-related words in commit message, just like the true positive samples. For example, as shown in Fig.~\ref{fig:rq5-2}, the non-security patch \href{https://github.com/torvalds/linux/commit/95609155d7fa08cc2e71d494acad39f72f0b4495}{\textit{9560915}} contains ``overflow check'' in commit message, but is not a security patch. Similarly, some of the false negative samples (i.e., misclassified as non-security patches) do not contain security-related words in commit message. For example, the security patch \href{https://github.com/FFmpeg/FFmpeg/commit/01e5e97026cf0b344abafca22b0336a2c58b2a33}{\textit{01e5e97}} with commit message `` mjpegbdec: Fix incorrect bitstream buffer size.'' does not contain any security-related words, but provides obvious security information in the commit message. } 
    \begin{figure}[t]
    \centering
    \includegraphics[width=8cm]{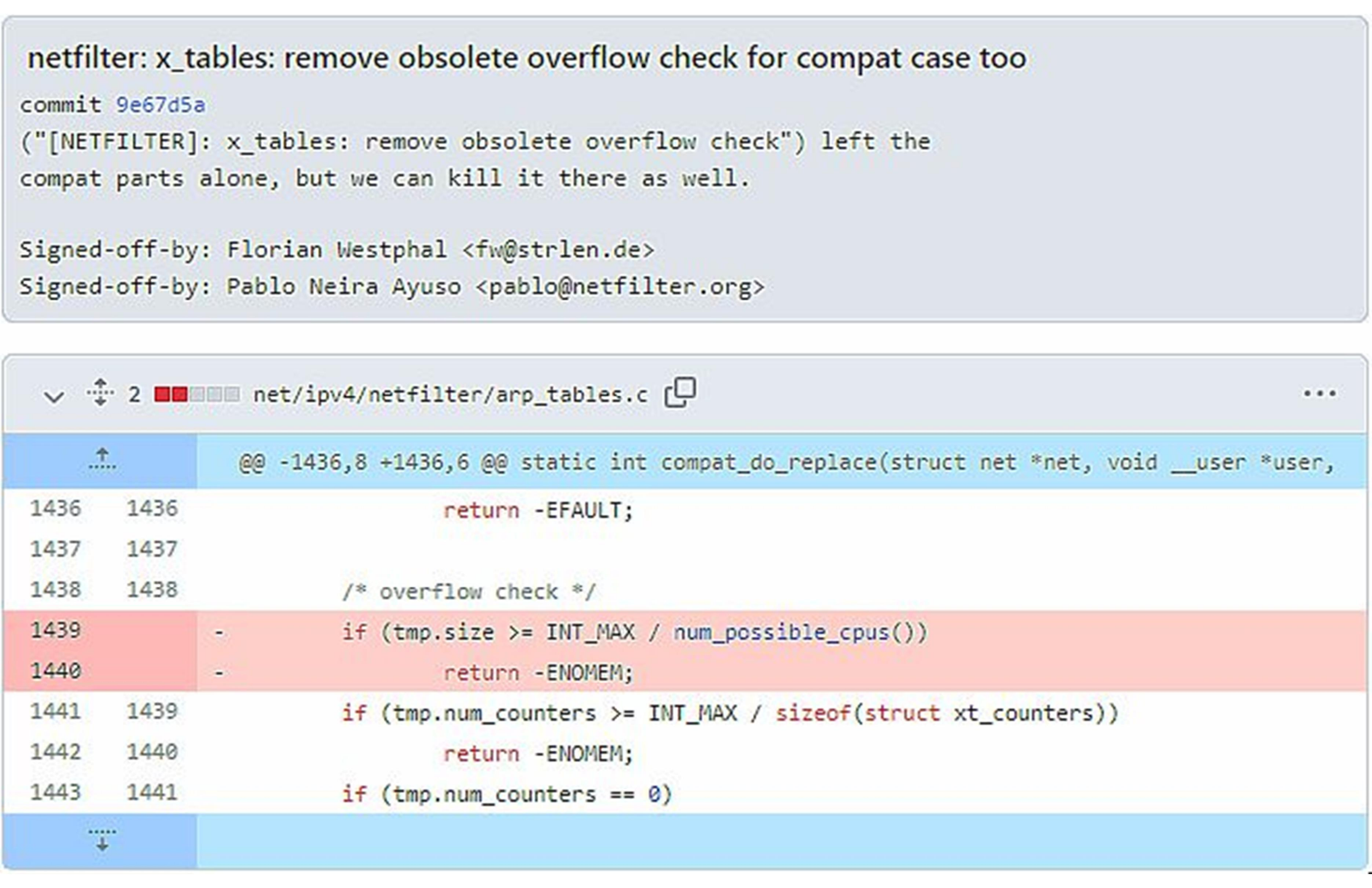}
    \caption{Commit \href{https://github.com/torvalds/linux/commit/95609155d7fa08cc2e71d494acad39f72f0b4495}{\textit{9560915}}. }
    \label{fig:rq5-2}
    \vspace{-4mm}
    \end{figure}
    
    
    \item \wbz{In addition, among the misclassified samples, we find that there are two mislabelled samples (i.e., commit \href{https://github.com/torvalds/linux/commit/695be0e7a24a8875c347437566f2c44ba673580b}{\textit{695be0e}} and commit \href{https://github.com/torvalds/linux/commit/3614364527daa870264f6dde77f02853cdecd02c}{\textit{3614364}}). Both are security patches, but labeled as non-security patches and \tool make the correct predictions for them. }
\end{itemize}

\wbz{From the phenomena, \tool seems to identify most security and non-security patches based on security-related words and security patterns, while also causing some samples to be misclassified. Therefore, we can infer that \tool pays close attention to security-related words in commit messages and security patterns in change code when identifying security patches. Furthermore, we also find that \tool can identify silent security patches which are mislabeled as non-security patches.}

\begin{tcolorbox}[breakable,width=\linewidth,boxrule=0pt,top=1pt, bottom=1pt, left=1pt,right=1pt, colback=gray!20,colframe=gray!20]
\textbf{Answer to RQ5:} \wbz{When identifying security patches, \tool pays close attention to security-related words in commit messages and security patterns in change code. }
\end{tcolorbox}

\section{Discussion}\label{sec:discussion}
In this section, we first discuss the impact of different numbers of paths (i.e., $k$) \wbz{and the ratio of within-changes paths and within-context paths} to the performance, \wbz{and then we conduct an online survey to investigate how developers evaluate the results generated by \tool. Finally, we discuss several potential limitations of \tool.}

\subsection{Choice of \textit{k}}\label{lb:choice_of_k}
\begin{figure}[t]
    \centering
    \includegraphics[scale=0.4]{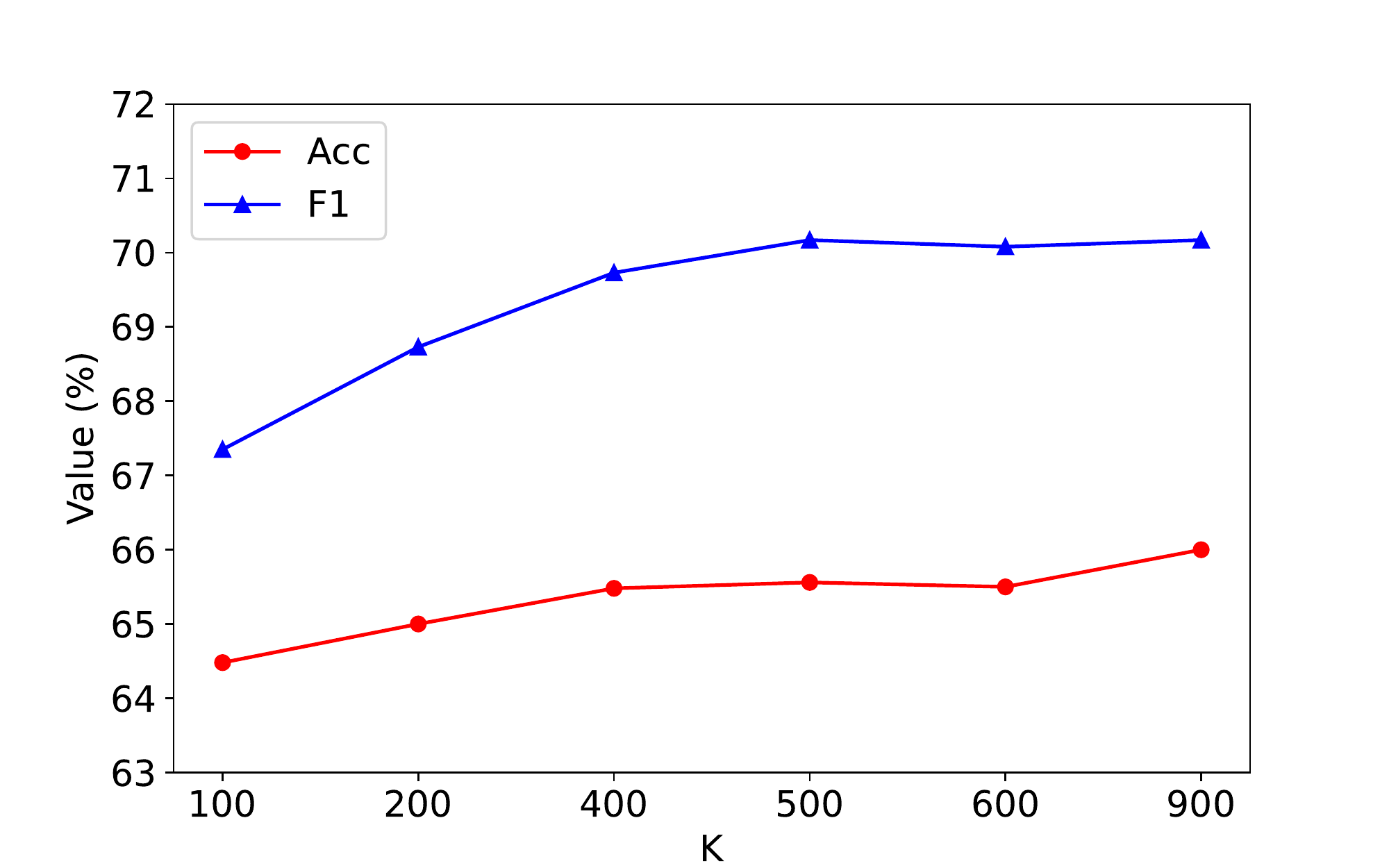}
    \caption{The impact of $k$ on accuracy and F1.}
    \label{fig:k}
\end{figure}

In Section~\ref{sec:settings}, we set the maximum number of AST paths $k$ to 500. We also tune $k$ within a range of 100 to 900 for AST-based code change encoder. Specifically, we turn off the commit message encoder and just use the code change encoder for the experiment to evaluate the effect of the number of paths on the code change encoder, and the other settings are the same as in Section~\ref{sec:settings} for a fair comparison. The experimental results are shown in Fig.~\ref{fig:k}.

From Fig.~\ref{fig:k}, we can observe that with increasing number of paths, accuracy and F1 are improved simultaneously, which intuitively reveals that the encoder can learn more effective information with increasing number of paths. After 500 paths, the accuracy and F1 do not improve significantly by further increasing the number of the paths. To balance performance and efficiency (i.e., more paths tend to raise the {computation} time), we set the AST number to 500 while representing the changed code in our experiments.

\subsection{The ratio of within-changes paths and within-context paths.}
\begin{figure}[t]
    \centering
    \includegraphics[scale=0.4]{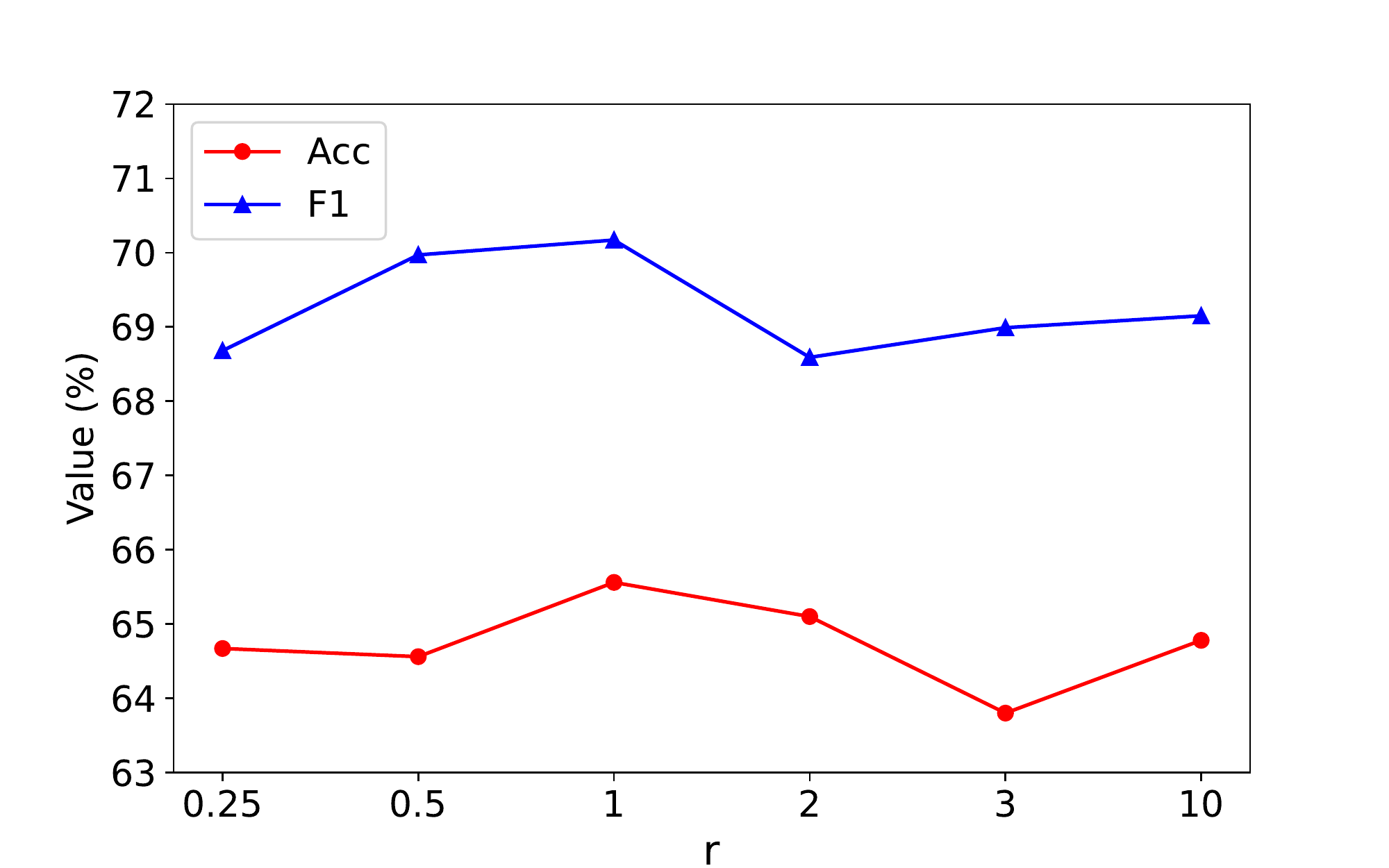}
    \caption{The impact of $r$ on accuracy and F1.}
    \label{fig:r}
\end{figure}

\wbz{Since the AST paths we extracted from source code include two parts: within-changes and within-context, the ratio of them may effect the final performance. Here, we define the ratio of within-changes paths to within-context paths as $r$. To study the effect of $r$ on the performance, we keep  the maximum number of AST paths $k$ to 500, and conduct experiments with different $r$ ranging from 0.25 to 10. The experiment settings are the same with Section~\ref{lb:choice_of_k}, and the results can be seen in Fig.~\ref{fig:r}. }

\wbz{We can see that when $r$ is equal to 1, the accuracy and F1 reach the peak. This means that the number of within-changes paths and within-context paths is the same, resulting in both types of information being fed into the model fairly. However, we can not guarantee that the model achieves the best performance when $r$ is equal to 1, since the relationship between $r$ and performance is not obvious. When using this model, $r$ needs to be tuned based on the dataset.}

\subsection{Online survey}
\wbz{We conduct an online survey to investigate how well developers trust the results generated by \tool. }

\noindent{\textbf{Dataset}.}
\wbz{We randomly select 10 samples from the test set and use the model prediction results to conduct the survey analysis.}

\noindent{\textbf{Participant Recruitment.}}
\wbz{We recruit 30 people from industrial companies and our universities to participate in the online survey. Among the participants, 16 of them come from industry and the rest are from academia. They are all software developers, from PhD students, post-doctoral researchers to software engineers. All of them are not coauthors of our work and use Github for software development with more than 3 years of development experience.}

\begin{figure}[t]
    \centering
    \includegraphics[scale=1.1]{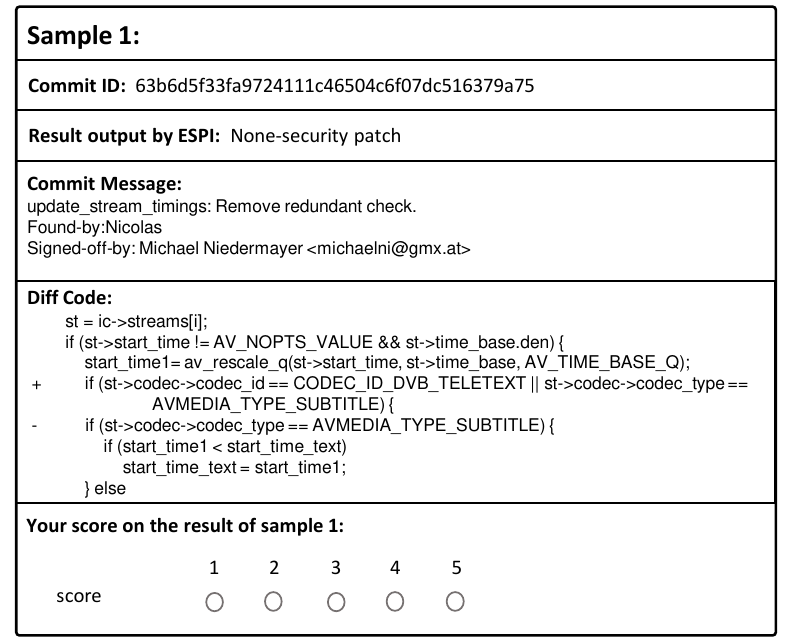}
    \caption{Part of the questions in the survey.}
    \label{tb:survey}
\end{figure}

\noindent{\textbf{Experiment Procedures.}}
\wbz{We start the online survey with a brief introduction. We explain to the participants that our task is to assess their level of trust in the results produced by \tool. Then the participants are required to provide their personal information relevant to the survey. To quantitatively measure how much they trust the results, we define the rating scale as 1 to 5 where a higher score means that the more confidence the participants have in the results. Commit messages and change code are presented to participants so that they can make their own predictions for each commit. After that, they are required to rate the confidence they have in the results of \tool by comparing them with their own prediction for the 10 test samples. They are required to complete the survey online, where Fig.~\ref{tb:survey} demonstrates part of the questions and the completed survey is available on {\url{https://forms.gle/m7SSxgqBYDk3mp9u9}}.}

\begin{figure}[t]
    \centering
    \includegraphics[scale=0.7]{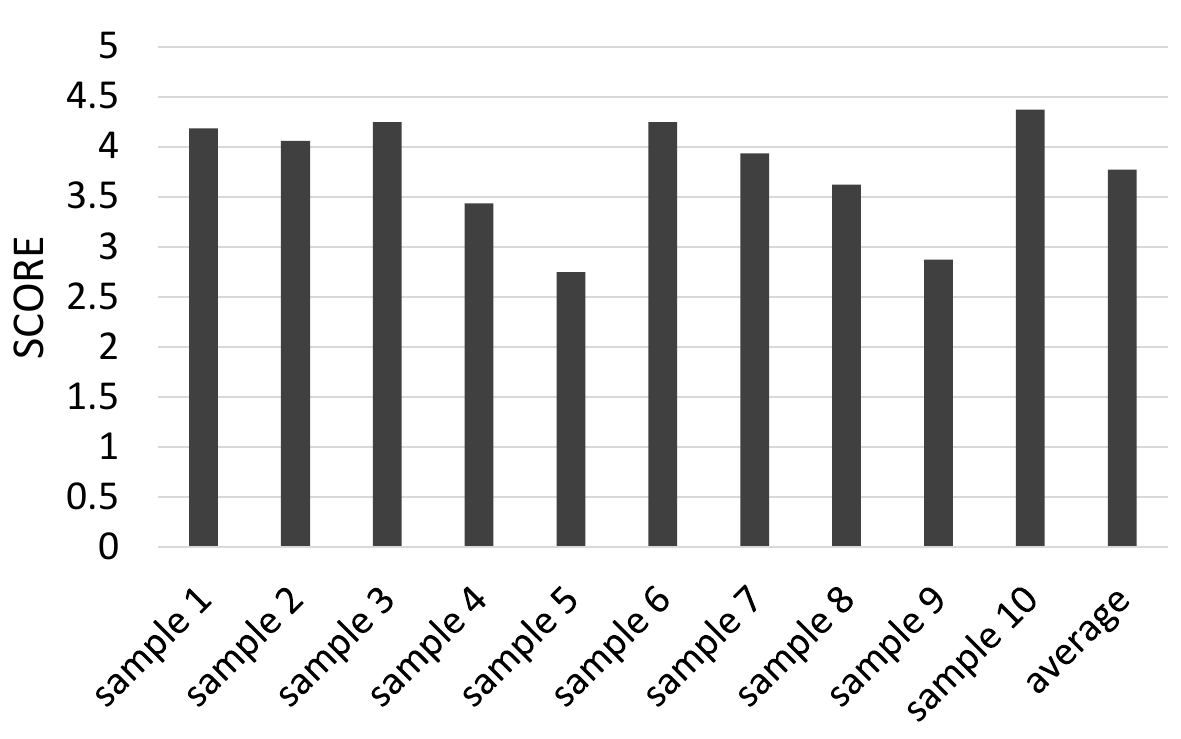}
    \caption{The average score of all samples in the survey.}
    \label{fig:survey}
\end{figure}

\noindent{\textbf{Survey Results}.}
\wbz{The average score of each sample is shown in Fig.~\ref{fig:survey}. We can see that the average scores of sample 1, 2, 3, 6 and 10 are above 4, which means that the results of \tool are in good agreement with the participants' own predictions. Except for sample 5 and 9, all other sample scores are greater than 3, which means that the results are acceptable for participants. Generally, the average score for all 10 samples is 3.78 which is better than Acceptable and close to Good.} 

\subsection{Limitations}
By comparing the results in Section~\ref{lb:exp1} and Section~\ref{sec:wild}, we find that the decrease in the real deployment is obvious. Although it is a common sense that deep learning-based techniques tend to have the poorer performance on the out-of-domain dataset, which is {different from the training data}, it actually hinders the model deployed to the real production environment and we also admit it is the limitation of our work. To mitigate this limitation, we propose some underlying solutions to improve the generalizability of the model.

\begin{itemize}[leftmargin=*]
\item{Improving the diversity of the training data.} It is a straightforward idea to expend the dataset. {Because the used data in our experiment is quite limited, including more commit data from more projects may improve the performance on unknown data. However, it is difficult in the real scenario since the cost of labeling a large scale of dataset is substantial.} Hence, an alternative option is to perform data augmentation~\cite{shorten2019survey} on the existing dataset to enrich the diversity. For example, we can perform some semantic-equivalent operators on the commit and {transform} it into different versions to improve the diversity of data for security patch identification. We leave the design for the semantic-equivalent augmentation on the commit as our future work. 
\item{Improving the generalization ability of the model.} In terms of the model design, although the generalization ability of the model is a long-term problem in deep learning community, recent advanced researches on the ensemble learning~\cite{kim2017domain, guo2018multi, clark2019don} have proved that the problem of the generalization capability can be mitigated. Hence, we could utilize ensemble learning, which combines multiple learners, to improve the generalization ability for security patch identification.  
\end{itemize}

\section{Related Work}\label{sec:related}
In this section, we briefly introduce related works about security patch identification, graph neural networks, and code representation techniques.
\subsection{Security Patch Identification}
A large amounts of silent security patches in the real world have not been noticed, causing them to be disclosed. Identifying these silent patches can help developers fix security bugs in their software in time and avoid the N-day attack. However, thousands of patches are submitted every day to Github, Bugzilla or other platforms to fix the bugs in their software. It is a great challenge to manually identify security patches in such a large number of patches. Therefore, automatic security patch identification becomes essential. Since the traditional static analysis approaches suffer from high false positive, people turn to learning-based approaches. In the early stages, conventional machine learning algorithms are applied to identify security patches. For example, Zhou et al.~\cite{zhou2017automated} propose to utilize a stacking model that assembles six individual classifiers for security patch identification. After that, deep learning-based approaches gradually became mainstream for this task due to their powerful capabilities. Wang et al.~\cite{wang2021patchrnn} propose PatchRNN, which ensembles two BiLSTM models to generate the feature vectors for each patch. Zhou et al.~\cite{zhou2021spi} propose SPI, which leverages LSTM and CNN together to learn the representation of each patch. Lozoya et al.~\cite{lozoya2021commit2vec} propose Commit2vec, which also employs BiLSTM to learn the representation from AST paths. We can also find that, in addition to commit message, people are trying to leverage the code information for security patches identification. In contrast to the approach in the preliminary stage ~\cite{zhou2017automated} which only considers commit messages as input, Commit2vec starts extracting AST paths from diff code as input, while PatchRNN and SPI both use commit messages along with code for security patch identification. Similar to PatchRNN and SPI, we leverage both commit message and code to identify security patches.

\subsection{Graph Neural Networks}
In our real world, there are many graph types of data, which is composed of objects and their relationships, such as social networks, molecules, and programs. Before the advent of GNN, due to the great progress of conventional neural networks such as CNN and LSTM, they have been widely applied in various fields such as image classification and natural language processing~\cite{wang2016cnn,lee2017going,li2014medical,liu2019text,nallapati2016abstractive}. However, people gradually find that these conventional neural networks can only operate on regular Euclidean data such as images and texts~\cite{bronstein2017geometric}. Therefore, graph neural networks are proposed to extend neural networks to non-Euclidean domains. In just a few years, many variants of graph neural networks have been proposed, which can be mainly categorized into four groups~\cite{wu2020comprehensive}: recurrent graph neural networks~\cite{li2015gated,dai2018learning}, convolutional graph neural networks~\cite{defferrard2016convolutional,kipf2016semi}, graph autoencoders~\cite{li2018learning,cao2016deep} and spatial-temporal graph neural networks~\cite{seo2018structured,jain2016structural}. In this paper, we apply GGNN, a type of recurrent graph neural network, to the dependency graph of commit messages for graph representation learning.

\subsection{Code Representation}
To represent code for machine learning, in a preliminary stage, the code is simply treated as a sequence, such as a sequence of tokens or a bag of words. RNN techniques such as LSTM and GRU are used to generate code representations. However, it was discovered that treating code as a sequence lost its structure information, since code are highly-structured data format. These program structures reveal the syntax and semantics of the code~\cite{siow2022learning, liu2020unified}. Therefore, many works are trying to use program structures for code representation such as ASTNN~\cite{astnn}, ATOM~\cite{liu2020atom}, Program Graph~\cite{allamanis2017learning}, Devign~\cite{zhou2019devign}, VulSniper~\cite{duan2019vulsniper} and HGNN~\cite{liu2021retrievalaugmented}. Specifically, Allamanis et al.~\cite{allamanis2017learning} uses a graph gated neural network and a program graph to learn function name and detect variable misuse. Devign~\cite{zhou2019devign} and Vulsniper~\cite{duan2019vulsniper} employ the code property graph to learn vulnerable functions. ASTNN~\cite{astnn} uses an AST representation to represent source code and perform source-code related tasks, such as clone detection and source code classification. Some other works have also attempted to use low-level programming representations such as LLVM. For example, Flow2Vec~\cite{flow2vec} employs an interprocedural value-flow graph and LLVM intermediate representation to learn a meaningful representation of the program. Compared with these works, we employ a Bi-LSTM to incorporate AST path information to learn the semantics of patches.
\section{Conclusion}\label{sec:conclusion}
In this paper, we propose a well-designed tool \tool for automated security patch identification, which fully utilizes the structure information in a commit for effective identification. By the AST-based code change encoder to extract the contextual AST paths about the changed code with BiLSTM to learn its representation and the graph-based commit message encoder to construct the dependency graph for the commit message with the graph neural network to learn the token relations {from} the message, \tool outperforms current state-of-the-art baselines significantly with {4.01\%} higher accuracy and {4.42\%} F1 score on the current dataset, increases 6.03\% accuracy and 7.38\% F1 in the real deployment environment.

\section{Acknowledgments}\label{sec:acknowledgments}
This research is partially supported by the National Research Foundation, Singapore under its the AI Singapore Programme (AISG2-RP-2020-019), the National Research Foundation, Prime Ministers Office, Singapore under its National Cybersecurity R\&D Program (Award No. NRF2018NCR-NCR005-0001), NRF Investigatorship NRF-NRFI06-2020-0001, the National Research Foundation through its National Satellite of Excellence in Trustworthy Software Systems (NSOE-TSS) project under the National Cybersecurity R\&D (NCR) Grant award no. NRF2018NCR-NSOE003-0001, the Ministry of Education, Singapore under its Academic Research Tier 3 (MOET32020-0004). Any opinions, findings and conclusions or recommendations expressed in this material are those of the author(s) and do not reflect the views of the Ministry of Education, Singapore.

%
\IEEEpeerreviewmaketitle

\bibliographystyle{IEEEtran}
\bibliography{ref}


\end{document}